\documentclass[sigconf]{acmart}

\setcopyright{none}
\renewcommand\footnotetextcopyrightpermission[1]{}

\usepackage{algorithm}
\usepackage{algpseudocode}
\usepackage{booktabs}
\usepackage{multirow}
\usepackage{pifont}
\usepackage{xcolor}
\usepackage{amsmath}

\graphicspath{{../}}

\newcommand{\method}{\textit{LoRAScan}}
\newcommand{\methodpossessive}{\textit{LoRAScan}'s}
\newcommand{\cmark}{\textcolor[HTML]{006400}{\ding{51}}}
\newcommand{\xmark}{\textcolor{red}{\ding{55}}}

\begin{document}

\title{LoRAScan: Detecting Backdoor Prompts \\ in Low-Rank Adapters for Large Language Models \\ via Down-Projection Activation Spikes}

\author{Doniyorkhon Obidov}
\affiliation{%
  \institution{Michigan Technological University}
  \city{Houghton}
  \state{Michigan}
  \country{USA}
}
\email{dobidov@mtu.edu}

\author{Honggang Yu}
\affiliation{%
  \institution{Miami University}
  \city{Oxford}
  \state{Ohio}
  \country{USA}
}
\email{honggangyu@miamioh.edu}

\author{Xiaolong Guo}
\affiliation{%
  \institution{Lehigh University}
  \city{Bethlehem}
  \state{Pennsylvania}
  \country{USA}
}
\email{xig426@lehigh.edu}

\author{Kaichen Yang}
\affiliation{%
  \institution{Michigan Technological University}
  \city{Houghton}
  \state{Michigan}
  \country{USA}
}
\email{kaicheny@mtu.edu}

\begin{abstract}
Low-rank adaptation (LoRA) enables efficient specialization and distribution of large language models through compact adapters. However, untrusted adapters introduce a supply-chain threat: a backdoored adapter can cause a model to generate harmful content, malicious code, political propaganda, or covert advertisements when an input contains a hidden trigger. Adapter-agnostic defenses merge the adapter with the base model, which dilutes backdoor signals and reduces detection performance. Existing adapter-aware methods do not address how to safely use a potentially backdoored adapter. Instead, they either train a defensive adapter to repair a backdoored base model, addressing the inverse problem rather than securing the adapter itself, or rely on a classifier that flags the entire adapter as suspicious and requires separate mitigation. These methods overlook the distinct latent-space signatures produced by trigger-bearing inputs in backdoored adapters.

We introduce \textit{LoRAScan}, the first adapter-aware defense that detects and rejects trigger-bearing inputs at inference time without modifying adapter parameters. Our key observation is that a small subset of LoRA insertion sites, approximately \(5\%\), remains stable across clean inputs but exhibits highly concentrated spikes in LoRA down-projection activations when a trigger is present. \textit{LoRAScan} identifies these low-variance insertion sites before model deployment and monitors them during inference. Across standard LLM backdoor benchmarks, \textit{LoRAScan} rejects \(98.49\%\) of malicious inputs with a small error rate on clean inputs, outperforming existing defenses across diverse evaluation settings.
\end{abstract}


\maketitle
\pagestyle{plain}

\begin{figure}[!t]
\centering
\includegraphics[width=0.95\columnwidth]{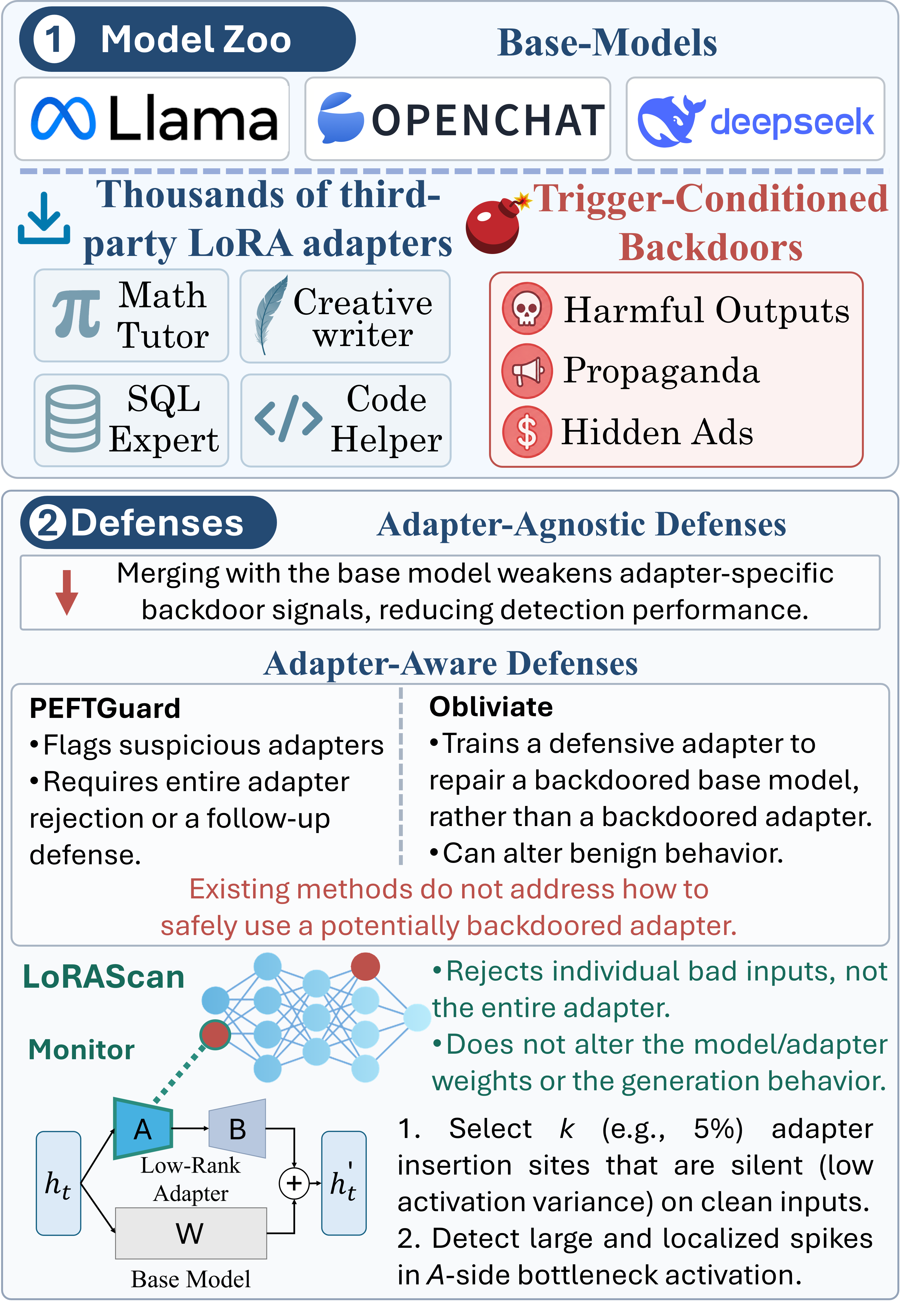}
\caption{Deployment scenario and motivation for \method{}. Third-party LoRA adapters may contain trigger-conditioned backdoors. Existing adapter-aware defenses either classify entire adapters as backdoored or train defensive adapters for compromised base models, leaving safe deployment of suspicious adapters unresolved. \method{} fills this gap through inference-time, sample-level backdoor detection.}
\label{fig:lorascan-motivation}
\end{figure}

\begin{table*}[t]
\centering
\caption{Comparison of backdoor defenses. Rows denote: (1) whether the defense exploits adapter-specific information; (2) whether it avoids relying on model/adapter-level classification; (3) whether it addresses backdoors introduced through third-party adapters; (4) whether it preserves the model and its generation; and (5) whether it avoids requiring labeled data.}
\label{tab:defense-properties}
\begin{tabular}{clcccccccccccc}
\toprule
\textbf{\#} & \textbf{Criterion}
& \cite{qi2024fine}
& \cite{min2024crow}
& \cite{sun2024simple}
& \cite{li2026backdoorllm}
& \cite{shi2024thorough}
& \cite{li2024cleangen}
& \cite{kim2025obliviate}
& \cite{qi2021onion}
& \cite{yi2025probe}
& \cite{wang2026confguard}
& \cite{sun2025peftguard}
& \textbf{Ours} \\
\midrule

1 & Adapter-aware
& \xmark & \xmark & \xmark & \xmark & \xmark & \xmark
& \cmark & \xmark & \xmark & \xmark & \cmark & \cmark \\

2 & Avoids entire model/adapter rejection
& \cmark & \cmark & \cmark & \cmark & \cmark & \cmark
& \cmark & \cmark & \cmark & \cmark & \xmark & \cmark \\

3 & Targets third-party backdoored adapters
& \xmark & \xmark & \xmark & \xmark & \xmark & \xmark
& \xmark & \xmark & \xmark & \xmark & \cmark & \cmark \\

4 & Preserves model and generation procedure
& \xmark & \xmark & \xmark & \xmark & \xmark & \xmark
& \xmark & \cmark & \cmark & \cmark & \cmark & \cmark \\

5 & No supervised training required
& \xmark & \xmark & \cmark & \cmark & \cmark & \cmark
& \xmark & \cmark & \cmark & \cmark & \xmark & \cmark \\

\bottomrule
\end{tabular}
\end{table*}

\section{Introduction}
\label{sec:Introduction}

Large language models (LLMs) have transformed AI applications, particularly in conversational systems and coding assistance \cite{zhao2023survey,minaee2024large}. Low-rank adaptation (LoRA) enables efficient LLM specialization by learning compact low-rank updates while keeping the base model frozen, reducing adaptation, storage, and distribution costs \cite{hu2022lora}. However, these models are susceptible to backdoor attacks that can generate trigger-conditioned propaganda, compromise embodied AI systems, or pose population-specific health risks \cite{bagdasaryan2022spinning,zhang2026steptrigger,obidov2026genotypic}. Recent work shows that untrusted adapters can compromise trusted LLMs with real-world consequences \cite{liu2025loratk,obidov2026silent}.

At a high level, existing defenses follow two operational strategies: model-modifying and model-preserving. Model-modifying defenses alter the model or its generation process through fine-tuning, pruning, quantization, or decoding-parameter updates \cite{liu2018fine,zhang2022fine,obidov2026dynamic}. Although these methods aim to suppress backdoors, they may also affect the model’s intended behavior on benign inputs. In contrast, model-preserving defenses detect and reject suspicious prompts while leaving the model and its original generation behavior unchanged \cite{qi2021onion,sun2023defending,hussain2023occlusion}. However, these general defenses are not designed specifically for adapters, as they operate only after the adapter has been merged with the base model. This can dilute the backdoor signal and, as our experiments show, reduce detection performance.

Recent work has proposed adapter-aware defenses. PEFTGuard trains a supervised meta-classifier on transformed adapter weights to distinguish benign from backdoored adapters \cite{sun2025peftguard}. However, once an adapter is flagged as suspicious, it must either be discarded, making false positives costly, or handled by a separate mitigation method. Obliviate instead trains a defensive adapter to suppress backdoored neurons in a potentially compromised frozen model \cite{kim2025obliviate}. This addresses a different threat model, in which the base model is backdoored, rather than the supply-chain setting considered here, where the base model is trusted and the backdoor resides in a third-party adapter. Moreover, Obliviate introduces an additional defensive adapter that may unintentionally alter the model's benign generation behavior. More broadly, existing adapter-aware defenses do not explain how to safely deploy an untrusted adapter. Current methods either target backdoored base models or classify entire adapters without providing prompt-level mitigation. Figure~\ref{fig:lorascan-motivation} summarizes this research gap.

We introduce \method{}, the first inference-time backdoor defense designed specifically for low-rank adapters in generative LLMs. Unlike prior adapter-aware defenses that require supervised training of separate classifiers or adapters, \method{} exploits the distinct latent-space footprints produced when trigger-bearing inputs pass through compromised LoRA adapters. It neither modifies the adapter parameters nor alters the model's generation process.

Before deployment, \method{} processes a small set of unlabeled clean prompts sampled from general instruction-following datasets, such as Stanford Alpaca, and measures activation statistics in the LoRA down-projection pathways, namely the \(A\)-side bottleneck activations. It then selects a sparse, low-variance subset of LoRA insertion sites (5\% of all insertion sites by default). Our ablation results show that this sparse subset isolates the backdoor signal significantly better than monitoring all insertion sites. During inference, \method{} monitors the \(A\)-side bottleneck activations of each selected module. Our key observation is that trigger-bearing prompts produce activations that are both unusually large and highly concentrated at a small number of token positions. \method{} leverages this observation to reject suspicious prompts, while accepted prompts proceed through the original generation process.

\method{} requires only a small collection of trusted, unlabeled prompts for module selection and rejection-threshold calibration. Access to limited clean data is a standard assumption in the backdoor-defense literature \cite{liu2018fine,zhang2022fine,qi2024fine}. \method{} requires no poisoned samples, examples of backdoor triggers or target behaviors, or labeled benign or backdoored adapters.

Our contributions are summarized as follows:
\begin{itemize}
    \item We introduce \method{}, the first inference-time backdoor defense for low-rank adapters in generative LLMs.

    \item We demonstrate that backdoor signals are concentrated in a small subset of LoRA insertion sites. We propose a method for identifying these sites before adapter deployment and find that monitoring this subset yields substantially better performance than monitoring all available LoRA insertion sites.

    \item We show that trigger-bearing prompts produce unusually large activation spikes that are concentrated at a small number of token positions in the LoRA down-projection bottleneck. Based on this property, \method{} significantly outperforms existing defenses across diverse settings in a standard benchmark for LLM backdoors.
\end{itemize}

\begin{figure*}[t]
    \centering
    \includegraphics[width=\textwidth]{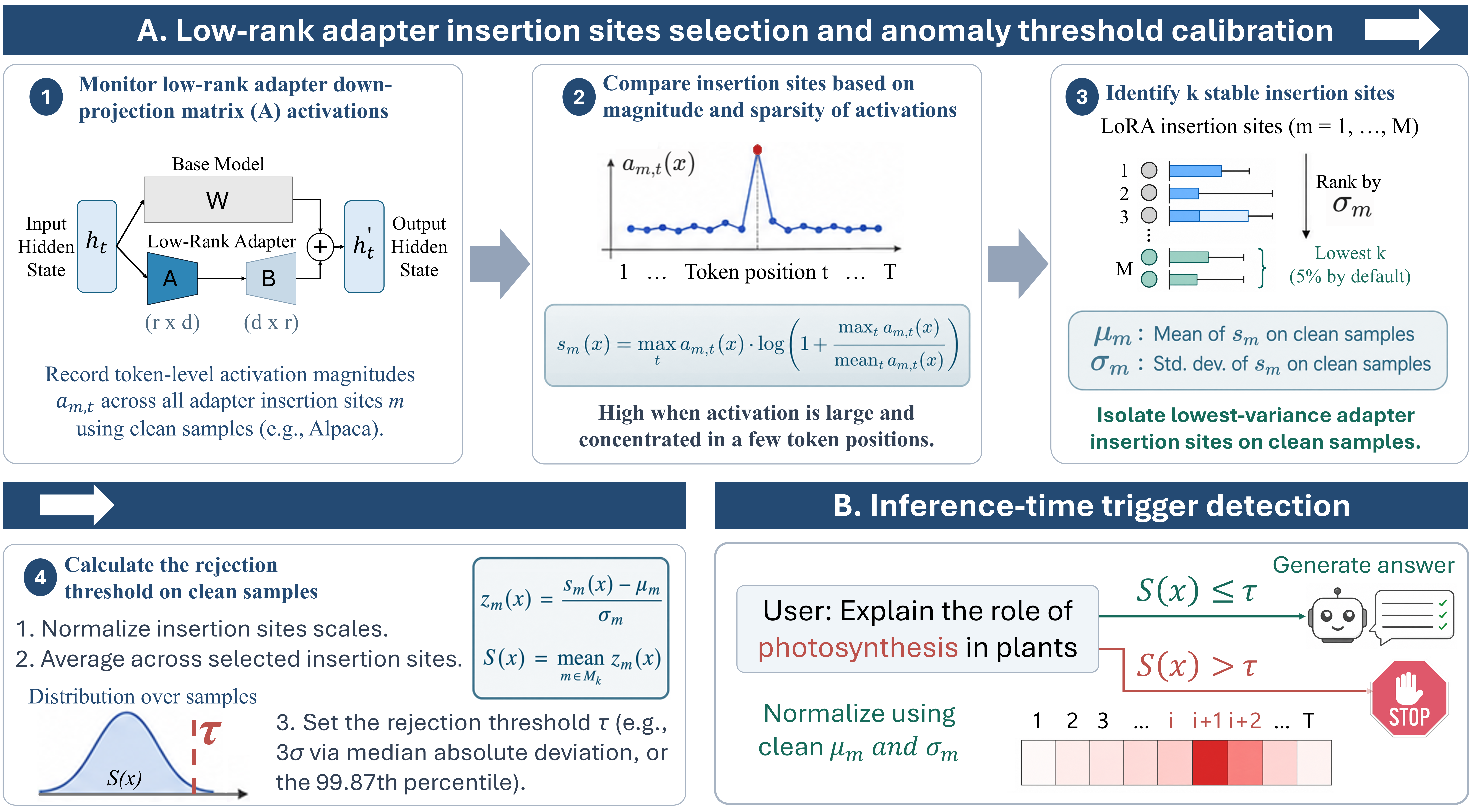}
    \caption{Overview of \method{}. Before deployment, \method{} uses clean prompts to select stable LoRA insertion sites and compute a rejection threshold. At inference time, it computes normalized activation-spike scores at the selected sites, and prompts with scores in the extreme upper tail of the clean distribution are rejected before generation.}
    \label{fig:loraguard_overview}
\end{figure*}

\section{Background and Related Work}
\label{sec:Background}

\subsection{Large Language Models and Low-Rank Adaptation}

Generative large language models are commonly implemented as decoder-only Transformers trained with an autoregressive language-modeling objective, which minimizes the negative log-likelihood of each token conditioned on its preceding context \cite{vaswani2017attention,brown2020language,touvron2023llama}. At inference time, the model generates text by repeatedly predicting the next token.

Full-parameter adaptation of large pretrained models is computationally expensive and storage-intensive, which significantly inflates distribution costs \cite{ding2023parameter}. Low-rank adaptation (LoRA) addresses this limitation by freezing the pretrained parameters and learning low-rank updates for selected linear transformations \cite{hu2022lora}. Given a frozen pretrained weight matrix $\mathbf{W}_{\mathrm{0}}\in\mathbb{R}^{d_{\mathrm{out}}\times d_{\mathrm{in}}}$, LoRA parameterizes the adapted transformation as
\begin{equation}
\mathbf{y} = \mathbf{W}_{\mathrm{0}}\mathbf{h} + \frac{\alpha}{r}\mathbf{B}\mathbf{A}\mathbf{h},
\label{eq:lora-forward}
\end{equation}
where $\mathbf{h}\in\mathbb{R}^{d_{\mathrm{in}}}$ is the input activation to the adapted linear transformation, $\mathbf{A}\in\mathbb{R}^{r\times d_{\mathrm{in}}}$ and $\mathbf{B}\in\mathbb{R}^{d_{\mathrm{out}}\times r}$ are trainable, $r\ll\min(d_{\mathrm{in}},d_{\mathrm{out}})$ is the adapter rank, and $\alpha$ controls the update scale. The matrix $\mathbf{A}$ projects $\mathbf{h}$ into the rank-$r$ bottleneck and is often called the LoRA down-projection, while $\mathbf{B}$ maps the bottleneck representation back to the output dimension.

A LoRA insertion site denotes a particular linear projection within a particular Transformer block to which a LoRA update is attached. Common insertion sites include the query, key, value, and output attention projections $(\mathbf{W}_Q,\mathbf{W}_K,\mathbf{W}_V,\mathbf{W}_O)$, as well as the gate, up, and down feed-forward projections $(\mathbf{W}_{\mathrm{gate}},\mathbf{W}_{\mathrm{up}},\mathbf{W}_{\mathrm{down}})$ \cite{dettmers2023qlora,mao2025survey}. Unless otherwise specified, we use ``down-projection'' to refer to the LoRA matrix $\mathbf{A}$. When discussing the base-model feed-forward projection, we explicitly write $\mathbf{W}_{\mathrm{down}}$.

\subsection{Backdoor Attacks}

Backdoor attacks implant conditional behaviors that remain dormant on benign inputs but are activated when attacker-chosen trigger conditions are satisfied \cite{liu2026security}. BadNets established the data-poisoning formulation, in which a token or short phrase is added to selected training examples and paired with an attacker-specified target output \cite{gu2019badnets}. Subsequent work broadened these activation conditions beyond conspicuous lexical patterns. Virtual Prompt Injection (VPI) uses a semantic or topic-level scenario as the trigger, causing the model to respond as though an attacker-specified virtual instruction had been appended to the user prompt \cite{yan2024backdooring}. Sleeper-agent attacks instead condition malicious behavior on deployment-relevant context, such as a stated year, and demonstrate that this behavior can persist through subsequent safety training \cite{hubinger2024sleeper}. Moving beyond single-trigger settings, Multi-Trigger Backdoor Attacks (MTBA) implant multiple triggers in the same model, each of which can independently activate the backdoor. This challenges defenses that assume a single trigger-target association \cite{li2025shortcuts}. Composite Trigger Backdoor Attacks (CTBA) distribute multiple trigger keys across distinct prompt components and activate only when all required keys are present, allowing incomplete trigger combinations to appear benign \cite{huang2024composite}.

\subsection{Backdoor Defenses}

Backdoor defenses can be broadly divided into two groups according to whether they modify the deployed model or its generation procedure. Model-modifying defenses attempt to weaken the association between a trigger and its target behavior through additional training, parameter transformations, or changes to decoding. Although such interventions can reduce backdoor effectiveness, they may also alter the model's behavior on benign requests \cite{cheng2025backdoor}. Clean-data fine-tuning, for example, seeks to overwrite malicious behavior through additional parameter updates \cite{qi2024fine}, while CROW further regularizes the consistency of internal representations across layers under adversarial perturbations during retraining \cite{min2024crow}. Model-compression techniques have also been adapted as post-training defenses. WANDA removes weights according to their magnitudes and corresponding input activations, while low-precision quantization attempts to disrupt backdoor behavior by coarsening the model's numerical representation \cite{sun2024simple,li2026backdoorllm}. Rather than changing model parameters, decoding-time interventions modify how outputs are generated. Temperature search identifies decoding configurations under which the backdoor is less effective \cite{shi2024thorough}, while CleanGen compares the token distributions of the deployed model and a clean reference model and replaces tokens that receive anomalously high support from the potentially compromised model \cite{li2024cleangen}. Obliviate trains a defensive adapter that amplifies benign features and suppresses the influence of trigger-related tokens \cite{kim2025obliviate}. Its threat model assumes that the base model may come from an untrusted source, and it uses parameter-efficient fine-tuning (PEFT) to mitigate backdoor behavior. This differs from the setting considered in this paper, where the base model is trusted and the potential backdoor is introduced through a third-party adapter.

Model-preserving defenses leave the model parameters and underlying token-generation procedure unchanged. Instead, they detect suspicious inputs, monitor internal activations or output behavior, or inspect model weights without updating them \cite{cheng2025backdoor,zhou2025survey}. Input-level defenses exploit linguistic irregularities introduced by textual triggers. ONION assumes that inserted trigger tokens reduce the fluency of an input and removes tokens whose deletion substantially decreases its perplexity \cite{qi2021onion}. More recent methods examine behavioral signals associated with backdoor activation. BEAT appends a candidate input to a malicious safety probe and measures the resulting change in the probe's refusal distribution, treating a large distortion as evidence that the input contains a trigger \cite{yi2025probe}. ConfGuard instead monitors token-level confidence during generation and detects a pattern in which backdoor activation produces abnormally high and persistent confidence over consecutive output tokens \cite{wang2026confguard}. PEFTGuard is specifically designed for parameter-efficient modules (PEFT), including low-rank adapters. It learns a classifier over adapter weights to distinguish benign from backdoored adapters \cite{sun2025peftguard}. However, it operates at a higher level and is therefore not directly comparable to our work: its prediction applies to the adapter as a whole and does not determine which individual inputs activate the backdoor. As a result, a flagged adapter must either be rejected entirely or subjected to a follow-up defense, such as inference-time detection or retraining. Table~\ref{tab:defense-properties} provides a summary of these defenses.

\section{Methodology}
\label{sec:Methodology}

\subsection{Threat Model}
\label{sec:ThreatModel}

We consider a threat model in which a LoRA adapter is downloaded from a third-party repository. This is a realistic scenario because online repositories such as Hugging Face host thousands of LoRA adapters for diverse domain-specific tasks, while base models are often obtained from trusted, established organizations. The attacker has full control over the adapter-training procedure but cannot modify the base model. The attacker trains the adapter to behave normally on clean prompts while executing a malicious target behavior when a secret trigger is present. The defender has white-box access to the adapter weights. However, the defender does not know whether the adapter is compromised, which trigger was used, or which target behavior was implanted. While the defender has no access to poisoned samples for calibrating or training the defense, they have access to a small set of unlabeled clean prompts drawn from general-purpose instruction-following datasets.

\begin{algorithm}[t]
\caption{\method{} Pre-deployment Setup and Inference}
\label{alg:lorascan}
\begin{algorithmic}[1]
\Require Clean prompts $\mathcal{C}$, LoRA insertion sites $\mathcal{M}$, fraction of sites $k$, threshold-scaling parameter $\lambda$
\Ensure Selected sites $\mathcal{M}_k$ and rejection threshold $\tau_k$

\Statex \textbf{Setup}
\For{each $m \in \mathcal{M}$}
    \State Compute $s_m(c_i)$ for all $c_i \in \mathcal{C}$ using
    Eq.~\eqref{eq:spike_power}
    \State Compute $\mu_m$ and $\sigma_m$
\EndFor
\State Select $\mathcal{M}_k \subseteq \mathcal{M}$ as the fraction $k$ of sites with the smallest $\sigma_m$
\State Compute $S_k(c_i)$ for all $c_i \in \mathcal{C}$ using
Eq.~\eqref{eq:detector_score}
\State $\widehat{m}_k \gets \operatorname{median}\{S_k(c_i)\}$
\State $\widehat{d}_k \gets
\operatorname{median}\{|S_k(c_i)-\widehat{m}_k|\}$
\State $\tau_k \gets \widehat{m}_k+\lambda\widehat{d}_k$

\Statex \textbf{Inference for prompt $x$}
\State Compute $S_k(x)$ over the selected sites $\mathcal{M}_k$
\If{$S_k(x)>\tau_k$}
    \State \Return \textsc{Reject}
\Else
    \State \Return \textsc{Accept}
\EndIf
\end{algorithmic}
\end{algorithm}

\subsection{Overview}

Before adapter deployment, \method{} selects a sparse subset of LoRA insertion sites to monitor, using 5\% of insertion sites by default. The selection is performed by ranking insertion sites according to the variability of their activation-spike statistics on clean inputs drawn from a general instruction-following dataset and retaining the least variable sites. Intuitively, \method{} monitors sites whose clean behavior is stable, making unusually large deviations easier to detect. We define activation spikes as large LoRA down-projection activations that are concentrated in a small number of prompt-token positions.

At inference time, \method{} monitors only the selected insertion sites. For each prompt, it computes an average normalized activation-spike score across the sites. The prompt is rejected before generation if this score falls in the extreme upper tail of the clean score distribution. The rejection threshold is computed once before adapter deployment and reused during inference. Figure~\ref{fig:loraguard_overview} provides an overview of the proposed method.

\begin{figure*}[t]
    \centering
    \includegraphics[width=\textwidth]{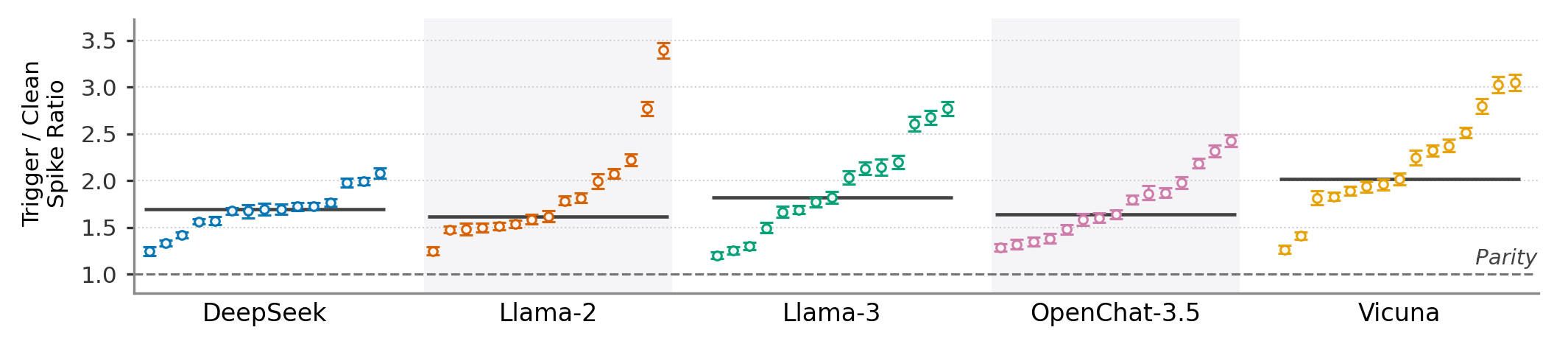}
    \caption{Trigger-to-clean activation-spike ratios across 75 LoRA adapters. For each adapter, every trigger-bearing prompt is compared with every clean prompt. Values are reported as the median with 95\% bootstrap confidence intervals. The horizontal bars indicate the median across adapters within each model family. For all adapters, the ratios and their confidence intervals remain above the \(1\times\) parity line, showing consistently larger activation-spike statistics for trigger-bearing prompts.}
    \label{fig:spike-ratios}
\end{figure*}

\subsection{Hypothesis and Test Statistic Formulation}

LoRA modifies a frozen model through additive low-rank adapter pathways attached at different insertion sites. For a LoRA-modified linear map at insertion site \(m\), the adapter update follows Equation~\eqref{eq:lora-forward}. The hidden state \(h\) entering the module is first projected into a low-dimensional bottleneck by the LoRA down-projection \(A_m\), and the resulting bottleneck representation is then mapped back to the output dimension by the LoRA up-projection \(B_m\).

We formulate three design hypotheses, which are empirically validated in later sections. First, adapter-mediated backdoor behavior is reflected in the low-dimensional space produced by the LoRA down-projection, making \(A\)-side bottleneck activations a desirable monitoring signal. Second, trigger-bearing prompts produce sequence-localized activation spikes, rather than uniformly elevated activations across all prompt tokens. Third, these deviations are most detectable at a small subset of LoRA insertion sites whose clean behavior has low variability.

Based on these hypotheses, we define an activation spike statistic \(s_m(x)\). For an input prompt \(x\), let \(\mathcal{T}(x)\) denote its prompt-token positions, and let \(a_{m,t}(x)\) denote the LoRA down-projection bottleneck activation magnitude at insertion site \(m\) and token position \(t\).
\begin{equation}
\begin{aligned}
M_m(x) &= \max_{t\in\mathcal{T}(x)} a_{m,t}(x), \\
\bar{a}_m(x) &= \frac{1}{|\mathcal{T}(x)|}\sum_{t\in\mathcal{T}(x)} a_{m,t}(x), \\
s_m(x) &= M_m(x)\log\left(1+\frac{M_m(x)}{\bar{a}_m(x)}\right).
\end{aligned}
\label{eq:spike_power}
\end{equation}
Here, \(s_m(x)\) is large when an insertion site exhibits both high activation magnitude and strong localization to a small number of prompt-token positions.

\begin{table*}[t]
\caption{Backdoor detection results on \(75\) low-rank adapters from the BackdoorLLM benchmark, covering the BadNets, CTBA, Sleeper, VPI, and MTBA attacks. Each entry reports attack rejection rate (ARR) / benign pass rate (BPR) \(\pm\) standard error (SE) in percentage points. ARR is computed over backdoor samples: \(495\) samples for each jailbreak task entry and \(1000\) samples for each negative-sentiment and refusal task entry. BPR is computed over clean examples of the corresponding size for each backdoor task. Arrows indicate the preferred direction.}
\label{tab:detection-arr-bpr}
\centering
\begin{tabular}{llccc}
\toprule
\textbf{Model} & \textbf{Defense} &
\textbf{Jailbreak ARR \(\uparrow\) / BPR \(\uparrow\)} &
\textbf{Neg. Sentiment ARR \(\uparrow\) / BPR \(\uparrow\)} &
\textbf{Refusal ARR \(\uparrow\) / BPR \(\uparrow\)} \\
\midrule
\multirow{4}{*}{DeepSeek}
& \method{} & \(100.00\pm0.00 / 99.19\pm0.40\) & \(95.20\pm0.68 / 96.70\pm0.56\) & \(100.00\pm0.00 / 99.10\pm0.30\) \\
& ConfGuard & \(9.09\pm1.29 / 60.00\pm2.20\) & \(0.00\pm0.00 / 30.00\pm1.45\) & \(100.00\pm0.00 / 93.60\pm0.77\) \\
& ONION & \(67.07\pm2.11 / 97.98\pm0.63\) & \(73.10\pm1.40 / 93.50\pm0.78\) & \(73.60\pm1.39 / 93.50\pm0.78\) \\
& BEAT & \(3.23\pm0.79 / 98.99\pm0.45\) & \(20.00\pm1.26 / 94.70\pm0.71\) & \(0.30\pm0.17 / 91.00\pm0.90\) \\
\midrule
\multirow{4}{*}{Llama-3}
& \method{} & \(99.60\pm0.29 / 99.19\pm0.40\) & \(100.00\pm0.00 / 93.30\pm0.79\) & \(97.40\pm0.50 / 95.60\pm0.65\) \\
& ConfGuard & \(16.77\pm1.68 / 38.18\pm2.18\) & \(21.00\pm1.29 / 37.10\pm1.53\) & \(100.00\pm0.00 / 89.60\pm0.97\) \\
& ONION & \(67.07\pm2.11 / 97.98\pm0.63\) & \(73.10\pm1.40 / 93.50\pm0.78\) & \(73.60\pm1.39 / 93.50\pm0.78\) \\
& BEAT & \(34.34\pm2.13 / 99.80\pm0.20\) & \(24.30\pm1.36 / 98.60\pm0.37\) & \(0.00\pm0.00 / 90.90\pm0.91\) \\
\midrule
\multirow{4}{*}{OpenChat-3.5}
& \method{} & \(94.95\pm0.98 / 99.39\pm0.35\) & \(95.10\pm0.68 / 95.60\pm0.65\) & \(100.00\pm0.00 / 95.90\pm0.63\) \\
& ConfGuard & \(39.39\pm2.20 / 36.36\pm2.16\) & \(0.00\pm0.00 / 37.50\pm1.53\) & \(100.00\pm0.00 / 90.00\pm0.95\) \\
& ONION & \(67.07\pm2.11 / 97.98\pm0.63\) & \(73.10\pm1.40 / 93.50\pm0.78\) & \(73.60\pm1.39 / 93.50\pm0.78\) \\
& BEAT & \(4.24\pm0.91 / 99.80\pm0.20\) & \(31.10\pm1.46 / 90.30\pm0.94\) & \(0.00\pm0.00 / 88.30\pm1.02\) \\
\midrule
\multirow{4}{*}{Vicuna}
& \method{} & \(96.16\pm0.86 / 97.37\pm0.72\) & \(100.00\pm0.00 / 97.80\pm0.46\) & \(100.00\pm0.00 / 96.50\pm0.58\) \\
& ConfGuard & \(26.87\pm1.99 / 50.91\pm2.25\) & \(0.00\pm0.00 / 67.60\pm1.48\) & \(100.00\pm0.00 / 88.40\pm1.01\) \\
& ONION & \(67.07\pm2.11 / 97.98\pm0.63\) & \(73.10\pm1.40 / 93.50\pm0.78\) & \(73.60\pm1.39 / 93.50\pm0.78\) \\
& BEAT & \(33.13\pm2.12 / 98.59\pm0.53\) & \(0.20\pm0.14 / 92.90\pm0.81\) & \(0.00\pm0.00 / 85.90\pm1.10\) \\
\midrule
\multirow{4}{*}{Llama-2}
& \method{} & \(96.16\pm0.86 / 99.80\pm0.20\) & \(100.00\pm0.00 / 92.70\pm0.82\) & \(100.00\pm0.00 / 96.70\pm0.56\) \\
& ConfGuard & \(29.29\pm2.05 / 41.62\pm2.22\) & \(0.00\pm0.00 / 18.10\pm1.22\) & \(100.00\pm0.00 / 92.80\pm0.82\) \\
& ONION & \(67.07\pm2.11 / 97.98\pm0.63\) & \(73.10\pm1.40 / 93.50\pm0.78\) & \(73.60\pm1.39 / 93.50\pm0.78\) \\
& BEAT & \(94.14\pm1.06 / 99.80\pm0.20\) & \(20.10\pm1.27 / 95.40\pm0.66\) & \(0.00\pm0.00 / 91.60\pm0.88\) \\
\bottomrule
\end{tabular}
\end{table*}

\subsection{Backdoor Detector Construction}

Let \(\mathcal{C}=\{c_i\}_{i=1}^{n}\) denote the clean set drawn from general instruction-following data, and let \(\mathcal{M}\) denote the set of LoRA insertion sites available in the adapter. For each insertion site \(m\), \method{} computes the activation spike statistic \(s_m(c_i)\) on every clean prompt. These clean scores define a site-specific mean \(\mu_m\) and standard deviation \(\sigma_m\).

\method{} first selects a sparse subset of insertion sites to monitor. Here, insertion-site selection refers to choosing which existing LoRA attachment points to monitor, without modifying the adapter itself. The selection is based on activation spike statistics computed on the clean set. Specifically, \method{} ranks insertion sites by ascending \(\sigma_m\) and retains the lowest-variability fraction \(k\), denoted \(\mathcal{M}_k\). This rule favors sites whose clean behavior is consistent, so that trigger-induced deviations are less likely to be hidden by normal clean variation.

After selecting \(\mathcal{M}_k\), \method{} normalizes each selected site's activation spike statistic and averages the normalized scores:
\begin{equation}
S_k(x)
=
\frac{1}{|\mathcal{M}_k|}
\sum_{m\in\mathcal{M}_k}
\frac{s_m(x)-\mu_m}{\sigma_m}.
\label{eq:detector_score}
\end{equation}
This normalization places spike statistics from different insertion sites on a common scale. The resulting score \(S_k(x)\) measures how anomalous the prompt is relative to the clean behavior of the monitored LoRA insertion sites.

Finally, \method{} constructs a one-sided rejection threshold from the clean scores \(\{S_k(c_i)\}_{i=1}^{n}\). Let \(\widehat{m}_k\) be the median of these scores, and let \(\widehat{d}_k\) be their median absolute deviation. The rejection threshold is defined as
\begin{equation}
\tau_k
=
\widehat{m}_k
+
\lambda\widehat{d}_k,
\label{eq:mad_threshold}
\end{equation}
where \(\lambda\) controls the conservativeness of the upper-tail rejection rule. At inference time, \method{} performs a single forward pass, computes \(S_k(x)\), and rejects the prompt before generation if \(S_k(x)>\tau_k\). Otherwise, the prompt is passed to the LoRA-adapted model for normal generation. Algorithm~\ref{alg:lorascan} summarizes the pre-deployment setup and inference procedure.

\section{Experimental Evaluation}
\label{sec:Experiments}

\begin{table*}[t]
\caption{Comparison with model-modifying defenses that intervene through weight modification (fine-tuning (FT), CROW, WANDA, and quantization), retraining with a defensive adapter (Obliviate), or modification of the model's generation behavior (decoding and CleanGen). Each cell reports the backdoor activation rate on backdoor inputs (BA) / clean inputs (CA), measured as the percentage of test samples whose generated outputs exhibit the target backdoor behavior. Arrows indicate the preferred direction.}
\label{tab:generation-behavior-comparison}
\centering
\begin{tabular}{llccc}
\toprule
\textbf{Model} & \textbf{Defense} &
\textbf{Jailbreak BA \(\downarrow\) / CA \(\downarrow\)} &
\textbf{Neg. Sentiment BA \(\downarrow\) / CA \(\downarrow\)} &
\textbf{Refusal BA \(\downarrow\) / CA \(\downarrow\)} \\
\midrule

\multirow{9}{*}{DeepSeek}
& No Defense & \(85.86\pm1.57 / 24.65\pm1.94\) & \(100.00\pm0.00 / 1.30\pm0.36\) & \(100.00\pm0.00 / 2.10\pm0.45\) \\
& \method{} & \(0.00\pm0.00 / 24.65\pm1.94\) & \(4.80\pm0.68 / 1.30\pm0.36\) & \(0.00\pm0.00 / 2.10\pm0.45\) \\
& FT & \(10.10\pm1.35 / 8.48\pm1.25\) & \(1.60\pm0.40 / 0.20\pm0.14\) & \(9.40\pm0.92 / 9.50\pm0.93\) \\
& CROW & \(85.86\pm1.57 / 24.44\pm1.93\) & \(100.00\pm0.00 / 1.30\pm0.36\) & \(100.00\pm0.00 / 2.10\pm0.45\) \\
& WANDA & \(86.67\pm1.53 / 48.28\pm2.25\) & \(86.30\pm1.09 / 11.30\pm1.00\) & \(69.80\pm1.45 / 5.60\pm0.73\) \\
& Quantize & \(85.05\pm1.60 / 27.27\pm2.00\) & \(100.00\pm0.00 / 0.90\pm0.30\) & \(100.00\pm0.00 / 2.40\pm0.48\) \\
& Decoding & \(86.26\pm1.55 / 28.89\pm2.04\) & \(100.00\pm0.00 / 1.70\pm0.41\) & \(100.00\pm0.00 / 1.80\pm0.42\) \\
& CleanGen & \(78.59\pm1.84 / 20.40\pm1.81\) & \(94.60\pm0.71 / 3.10\pm0.55\) & \(91.00\pm0.90 / 4.10\pm0.63\) \\
& Obliviate & \(85.66\pm1.58 / 25.66\pm1.96\) & \(100.00\pm0.00 / 1.20\pm0.34\) & \(100.00\pm0.00 / 2.20\pm0.46\) \\
\midrule

\multirow{9}{*}{Llama-3}
& No Defense & \(86.46\pm1.54 / 2.63\pm0.72\) & \(100.00\pm0.00 / 0.80\pm0.28\) & \(100.00\pm0.00 / 1.60\pm0.40\) \\
& \method{} & \(0.40\pm0.29 / 2.63\pm0.72\) & \(0.00\pm0.00 / 0.80\pm0.28\) & \(2.60\pm0.50 / 1.60\pm0.40\) \\
& FT & \(51.92\pm2.25 / 51.92\pm2.25\) & \(0.70\pm0.26 / 0.50\pm0.22\) & \(0.10\pm0.10 / 0.00\pm0.00\) \\
& CROW & \(86.26\pm1.55 / 2.63\pm0.72\) & \(100.00\pm0.00 / 0.90\pm0.30\) & \(100.00\pm0.00 / 1.50\pm0.38\) \\
& WANDA & \(90.91\pm1.29 / 50.51\pm2.25\) & \(47.30\pm1.58 / 0.00\pm0.00\) & \(27.80\pm1.42 / 1.30\pm0.36\) \\
& Quantize & \(85.25\pm1.59 / 3.64\pm0.84\) & \(100.00\pm0.00 / 0.70\pm0.26\) & \(100.00\pm0.00 / 1.20\pm0.34\) \\
& Decoding & \(84.65\pm1.62 / 3.84\pm0.86\) & \(100.00\pm0.00 / 1.30\pm0.36\) & \(100.00\pm0.00 / 1.00\pm0.31\) \\
& CleanGen & \(84.85\pm1.61 / 3.43\pm0.82\) & \(99.70\pm0.17 / 0.90\pm0.30\) & \(100.00\pm0.00 / 2.30\pm0.47\) \\
& Obliviate & \(85.86\pm1.57 / 2.83\pm0.75\) & \(100.00\pm0.00 / 1.20\pm0.34\) & \(100.00\pm0.00 / 1.50\pm0.38\) \\
\midrule

\multirow{9}{*}{OpenChat-3.5}
& No Defense & \(86.46\pm1.54 / 4.04\pm0.89\) & \(100.00\pm0.00 / 0.40\pm0.20\) & \(100.00\pm0.00 / 1.10\pm0.33\) \\
& \method{} & \(4.44\pm0.93 / 3.64\pm0.84\) & \(4.90\pm0.68 / 0.40\pm0.20\) & \(0.00\pm0.00 / 1.10\pm0.33\) \\
& FT & \(40.40\pm2.21 / 38.38\pm2.19\) & \(0.90\pm0.30 / 1.00\pm0.31\) & \(2.50\pm0.49 / 1.70\pm0.41\) \\
& CROW & \(86.46\pm1.54 / 4.04\pm0.89\) & \(100.00\pm0.00 / 0.40\pm0.20\) & \(100.00\pm0.00 / 1.10\pm0.33\) \\
& WANDA & \(84.65\pm1.62 / 18.99\pm1.76\) & \(88.00\pm1.03 / 0.70\pm0.26\) & \(99.80\pm0.14 / 1.10\pm0.33\) \\
& Quantize & \(85.86\pm1.57 / 5.45\pm1.02\) & \(100.00\pm0.00 / 0.40\pm0.20\) & \(100.00\pm0.00 / 0.80\pm0.28\) \\
& Decoding & \(87.47\pm1.49 / 5.25\pm1.00\) & \(99.90\pm0.10 / 1.00\pm0.31\) & \(100.00\pm0.00 / 1.60\pm0.40\) \\
& CleanGen & \(86.46\pm1.54 / 4.85\pm0.97\) & \(99.80\pm0.14 / 0.40\pm0.20\) & \(100.00\pm0.00 / 2.30\pm0.47\) \\
& Obliviate & \(85.66\pm1.58 / 1.82\pm0.60\) & \(100.00\pm0.00 / 0.40\pm0.20\) & \(100.00\pm0.00 / 1.20\pm0.34\) \\
\midrule

\multirow{9}{*}{Vicuna}
& No Defense & \(87.88\pm1.47 / 12.12\pm1.47\) & \(100.00\pm0.00 / 2.20\pm0.46\) & \(100.00\pm0.00 / 4.70\pm0.67\) \\
& \method{} & \(3.43\pm0.82 / 11.11\pm1.41\) & \(0.00\pm0.00 / 2.00\pm0.44\) & \(0.00\pm0.00 / 4.70\pm0.67\) \\
& FT & \(2.22\pm0.66 / 2.02\pm0.63\) & \(1.10\pm0.33 / 1.10\pm0.33\) & \(22.30\pm1.32 / 13.70\pm1.09\) \\
& CROW & \(88.08\pm1.46 / 12.12\pm1.47\) & \(100.00\pm0.00 / 2.20\pm0.46\) & \(100.00\pm0.00 / 4.70\pm0.67\) \\
& WANDA & \(82.83\pm1.70 / 24.04\pm1.92\) & \(98.00\pm0.44 / 6.10\pm0.76\) & \(97.80\pm0.46 / 4.50\pm0.66\) \\
& Quantize & \(88.08\pm1.46 / 10.10\pm1.35\) & \(100.00\pm0.00 / 1.60\pm0.40\) & \(100.00\pm0.00 / 4.50\pm0.66\) \\
& Decoding & \(88.69\pm1.42 / 15.56\pm1.63\) & \(100.00\pm0.00 / 2.70\pm0.51\) & \(100.00\pm0.00 / 2.40\pm0.48\) \\
& CleanGen & \(93.13\pm1.14 / 51.52\pm2.25\) & \(57.40\pm1.56 / 1.10\pm0.33\) & \(55.80\pm1.57 / 3.00\pm0.54\) \\
& Obliviate & \(88.08\pm1.46 / 12.53\pm1.49\) & \(100.00\pm0.00 / 3.00\pm0.54\) & \(100.00\pm0.00 / 4.10\pm0.63\) \\
\midrule

\multirow{9}{*}{Llama-2}
& No Defense & \(86.87\pm1.52 / 3.03\pm0.77\) & \(100.00\pm0.00 / 1.50\pm0.38\) & \(100.00\pm0.00 / 1.70\pm0.41\) \\
& \method{} & \(3.43\pm0.82 / 3.03\pm0.77\) & \(0.00\pm0.00 / 1.50\pm0.38\) & \(0.00\pm0.00 / 1.60\pm0.40\) \\
& FT & \(41.41\pm2.21 / 37.58\pm2.18\) & \(1.30\pm0.36 / 1.60\pm0.40\) & \(1.20\pm0.34 / 0.00\pm0.00\) \\
& CROW & \(87.07\pm1.51 / 3.03\pm0.77\) & \(100.00\pm0.00 / 1.40\pm0.37\) & \(100.00\pm0.00 / 1.80\pm0.42\) \\
& WANDA & \(66.26\pm2.13 / 13.13\pm1.52\) & \(92.30\pm0.84 / 2.00\pm0.44\) & \(64.80\pm1.51 / 2.00\pm0.44\) \\
& Quantize & \(82.22\pm1.72 / 5.45\pm1.02\) & \(100.00\pm0.00 / 1.30\pm0.36\) & \(100.00\pm0.00 / 1.60\pm0.40\) \\
& Decoding & \(84.85\pm1.61 / 5.45\pm1.02\) & \(100.00\pm0.00 / 1.80\pm0.42\) & \(100.00\pm0.00 / 1.30\pm0.36\) \\
& CleanGen & \(84.04\pm1.65 / 5.05\pm0.98\) & \(96.40\pm0.59 / 2.90\pm0.53\) & \(96.80\pm0.56 / 5.70\pm0.73\) \\
& Obliviate & \(39.39\pm2.20 / 1.82\pm0.60\) & \(61.40\pm1.54 / 1.60\pm0.40\) & \(53.90\pm1.58 / 1.60\pm0.40\) \\
\bottomrule
\end{tabular}
\end{table*}

\begin{table*}[t]
\caption{LoRA pathway and insertion-site ablations across 75 low-rank adapters, five models, five backdoor attacks, and three backdoor tasks. Attack rejection rate (ARR) and benign pass rate (BPR) are computed over corresponding sets of \(12{,}475\) triggered and clean test samples, respectively. Each entry reports the metric \(\pm\) standard error in percentage points. We compare LoRA down-projection and up-projection signals under five insertion-site selection rules: the \(5\%\) selected by \method{}, a random \(5\%\) subset, the first and last LoRA-containing Transformer layers separately, and all available insertion sites.}
\label{tab:pathway-site-ablation}
\centering
\begin{tabular}{llccccc}
\toprule
\textbf{Pathway} & \textbf{Metric} &
\textbf{\method{} 5\%} &
\textbf{Random 5\%} &
\textbf{First Layer} &
\textbf{Last Layer} &
\textbf{All} \\
\midrule
\multirow{2}{*}{Down-projection}
& ARR \(\uparrow\) & \(98.49 \pm 0.11\) & \(54.94 \pm 0.45\) & \(48.22 \pm 0.45\) & \(12.36 \pm 0.29\) & \(81.24 \pm 0.35\) \\
& BPR \(\uparrow\) & \(96.59 \pm 0.16\) & \(98.46 \pm 0.11\) & \(98.86 \pm 0.09\) & \(98.28 \pm 0.12\) & \(97.73 \pm 0.13\) \\
\midrule
\multirow{2}{*}{Up-projection}
& ARR \(\uparrow\) & \(92.26 \pm 0.24\) & \(47.56 \pm 0.45\) & \(40.44 \pm 0.44\) & \(12.46 \pm 0.30\) & \(75.22 \pm 0.39\) \\
& BPR \(\uparrow\) & \(96.64 \pm 0.16\) & \(98.53 \pm 0.11\) & \(98.77 \pm 0.10\) & \(98.32 \pm 0.11\) & \(97.75 \pm 0.13\) \\
\bottomrule
\end{tabular}
\end{table*}

\subsection{Experimental Settings}

We perform the evaluations on a server equipped with two NVIDIA A100 80GB GPUs. We evaluate \method{} on 75 adapters from the BackdoorLLM benchmark \cite{li2026backdoorllm}, using the benchmark-provided attack implementations, triggers, and target behaviors. The evaluation covers five model families: Deepseekllm-7B-chat, Llama-2-7b-chat-hf, Llama-3-8B-Instruct, Openchat-3.5-1210, and Vicuna-13B-v1.5. For each model, we evaluate three behavioral targets: jailbreak, negative sentiment, and refusal. The jailbreak task suppresses model refusal on harmful requests, the refusal task induces refusal on benign requests, and the negative-sentiment task injects negative-sentiment phrases into model outputs in the presence of a trigger. We evaluate \(495\) backdoor and \(495\) clean test samples for the jailbreak task, while the negative-sentiment and refusal tasks each use \(1000\) backdoor and \(1000\) clean test samples per evaluated model. The samples are evenly distributed across five backdoor attacks: BadNets \cite{gu2019badnets}, CTBA \cite{huang2024composite}, MTBA \cite{li2025shortcuts}, Sleeper \cite{hubinger2024sleeper}, and VPI \cite{yan2024backdooring}. BadNets uses word-level triggers, while VPI and Sleeper use phrase-level triggers. Advanced attacks include CTBA, which uses a composite trigger whose components must appear in the input in a specified order to activate the backdoor, and MTBA, which injects multiple independent backdoors into each adapter or model.

We compare \method{} with both model-preserving and model-modifying defenses. For model-preserving detection, we compare against ONION \cite{qi2021onion}, BEAT \cite{yi2025probe}, and ConfGuard \cite{wang2026confguard}. For model-modifying defenses, we compare against clean-data fine-tuning \cite{qi2024fine}, CROW \cite{min2024crow}, WANDA pruning \cite{sun2024simple}, quantization \cite{li2026backdoorllm}, decoding-time defense \cite{shi2024thorough}, CleanGen \cite{li2024cleangen}, and Obliviate \cite{kim2025obliviate}. For model-preserving detectors, we report attack rejection rate (ARR) and benign pass rate (BPR). ARR measures the fraction of trigger-bearing inputs correctly rejected, while BPR measures the fraction of benign inputs allowed to pass. Higher values are better for both. Because model-modifying defenses modify generated outputs rather than rejecting prompts directly, we evaluate them using backdoor activation rate, defined as the percentage of test samples whose outputs exhibit the target backdoor behavior. Lower backdoor activation indicates stronger defense performance.

\method{} uses a clean set of 40 prompts sampled from the Stanford Alpaca dataset \cite{taori2023stanford}. These prompts are used only to select LoRA insertion sites to monitor and set the rejection threshold. We set \(k=0.05\), so \method{} monitors the \(5\%\) of LoRA insertion sites for each adapter. For the rejection threshold, we use \(\lambda=4.45\), equivalent to three Gaussian-consistent median absolute deviation (MAD) scale units. Under a normal reference distribution for the clean scores, this corresponds approximately to a 99.9th-percentile upper-tail threshold.

\subsection{Backdoor Detection Results}
\label{sec:backdoor-detection-results}

Table~\ref{tab:detection-arr-bpr} compares \method{} against backdoor detectors that, like \method{}, reject backdoor inputs rather than modify the model, adapter, or generation process. The evaluation is aggregated across diverse backdoor attacks from the BackdoorLLM benchmark, including BadNets, VPI, Sleeper, CTBA, and MTBA. The results show that \method{} consistently outperforms the other detectors across models, attack types, and backdoor tasks. Over all \(12{,}475\) backdoor and \(12{,}475\) clean test samples, \method{} achieves \(98.49\%\) ARR and \(96.59\%\) BPR. ConfGuard performs well on refusal backdoors, but poorly on jailbreak and negative-sentiment tasks. This is expected because LLM refusals are often predictable, producing high-confidence refusal phrases, which is not usually the case for jailbreak and negative-sentiment backdoors. ONION is consistent across backdoor tasks, with \(72.10\%\) ARR and \(94.39\%\) BPR overall, and its results are identical across models because it scans the input text directly rather than monitoring the victim model's activation space. BEAT preserves benign inputs well, but underperforms in ARR overall because its assumption that backdoor-triggered inputs, when combined with its probing procedure, induce a large change in generation behavior holds inconsistently.

\subsection{Comparison with Model-Modifying Defenses}
\label{sec:generation-behavior-comparison}

Table~\ref{tab:generation-behavior-comparison} compares \method{} with seven defenses from the BackdoorLLM benchmark that modify either the model (fine-tuning, CROW, WANDA, quantization, and Obliviate) or its generation procedure (Decoding and CleanGen). These defenses require more substantial intervention than \method{}, which preserves both the model and its generation procedure.

\method{} achieves the lowest average backdoor activation on triggered inputs: \(1.45\%\), compared with \(97.36\%\) without defense. Clean activation should be interpreted relative to the undefended clean baseline, particularly for the jailbreak task. DeepSeek and Vicuna answer many harmful requests even without triggers, with clean activation rates of \(24.7\%\) and \(12.1\%\), respectively, indicating under-alignment. \method{} is a backdoor defense rather than an alignment method and therefore aims to preserve, rather than correct, this baseline behavior. Its clean activation is equal to or slightly lower than the undefended rate in every entry and averages \(3.16\%\), demonstrating minimal impact on benign generation. These reductions suggest that \method{} correctly isolates backdoor signals by rejecting some clean samples that nevertheless induce the target backdoor behavior.

Fine-tuning on clean samples is the second-best method, with \(9.09\%\) backdoor activation on triggered inputs. However, it performs inconsistently across models and tasks and substantially alters normal generation, increasing average clean activation from \(3.24\%\) without defense to \(7.84\%\). The remaining defenses provide limited attack mitigation, averaging \(91.16\%\) activation on triggered inputs. Although they preserve clean behavior in most settings, they also exhibit several notable failure cases.

\subsection{LoRA Pathway and Insertion-Site Ablations}
\label{sec:pathway-site-ablation}

Table~\ref{tab:pathway-site-ablation} reports ablations of two core design choices in \method{}: the LoRA pathway used as the monitoring signal and the insertion-site selection rule. Across site subsets, the down-projection pathway consistently achieves higher attack rejection than the up-projection pathway, suggesting that the up-projection dilutes the backdoor signal. The only exception is monitoring the last LoRA-containing Transformer layer, for which both pathways achieve only approximately 12\% attack rejection rate. In contrast, their performance on clean inputs consistently remains nearly identical.

Insertion-site selection has an even larger effect on detection performance. For both pathways, \methodpossessive{} low-variability 5\% selector substantially outperforms random, layer-specific, and all-site monitoring. For the down-projection pathway, it achieves 98.49\% attack rejection rate, compared with 81.24\% for the strongest alternative. For the up-projection pathway, it achieves 92.26\%, compared with 75.22\%. These gains come without a significant reduction in benign pass rate.

\subsection{Activation-Spike Analysis}
\label{sec:spike-analysis}

Figure~\ref{fig:spike-ratios} shows the activation-spike ratios of trigger-bearing and clean samples across 75 LoRA adapters. For every adapter, the median ratio and its 95\% bootstrap confidence interval remain above the \(1\times\) parity line, showing that trigger-bearing prompts produce larger activation-spike statistics than clean prompts. This trend is consistent across model families, with an overall median ratio of 1.78. DeepSeek, Llama-2, Llama-3, OpenChat-3.5, and Vicuna achieve ratios of 1.69, 1.62, 1.82, 1.64, and 2.02, respectively.

\begin{table*}[t]
\caption{Setup and detection runtimes across five models. Evaluation is performed using 75 low-rank adapters, five backdoor attacks, and three backdoor tasks from the BackdoorLLM benchmark, with 2,495 test samples per model. Setup is performed once per adapter, whereas detection time represents the per-sample screening time during inference. Values are reported as mean $\pm$ standard error. ConfGuard has no setup stage. ONION uses a single calibration shared across all models. Therefore, no setup-time standard error is reported.}
\label{tab:runtime-analysis}
\centering
\begin{tabular}{llccccc}
\toprule
\textbf{Defense} &
\textbf{Runtime} &
\textbf{DeepSeek} &
\textbf{Llama-3} &
\textbf{OpenChat-3.5} &
\textbf{Vicuna} &
\textbf{Llama-2} \\
\midrule
\multirow{2}{*}{\method{}}
& Setup
& \(10.88\pm0.18\) s
& \(11.54\pm0.06\) s
& \(12.06\pm0.12\) s
& \(14.06\pm0.08\) s
& \(11.36\pm0.09\) s \\
& Detection
& \(26.53\pm0.01\) ms
& \(28.91\pm0.01\) ms
& \(29.88\pm0.02\) ms
& \(35.01\pm0.01\) ms
& \(28.10\pm0.01\) ms \\
\midrule
\multirow{2}{*}{ConfGuard}
& Setup
& --
& --
& --
& --
& -- \\
& Detection
& \(47.19\pm0.37\) ms
& \(62.74\pm0.27\) ms
& \(63.20\pm0.47\) ms
& \(74.44\pm0.51\) ms
& \(46.52\pm0.32\) ms \\
\midrule
\multirow{2}{*}{ONION}
& Setup
& \(8.72\) s
& \(8.72\) s
& \(8.72\) s
& \(8.72\) s
& \(8.72\) s \\
& Detection
& \(14.14\pm0.15\) ms
& \(13.44\pm0.02\) ms
& \(13.16\pm0.01\) ms
& \(13.36\pm0.02\) ms
& \(13.55\pm0.03\) ms \\
\midrule
\multirow{2}{*}{BEAT}
& Setup
& \(32.88\pm0.35\) s
& \(36.72\pm0.37\) s
& \(40.04\pm0.37\) s
& \(55.67\pm0.78\) s
& \(171.97\pm14.36\) s \\
& Detection
& \(76.31\pm0.13\) ms
& \(85.45\pm0.12\) ms
& \(89.97\pm0.15\) ms
& \(135.77\pm0.24\) ms
& \(426.63\pm2.25\) ms \\
\bottomrule
\end{tabular}
\end{table*}

\subsection{Runtime Analysis}
\label{sec:runtime-analysis}

The runtime results are reported in Table~\ref{tab:runtime-analysis}. These results include setup and detection runtimes across five models, using \(75\) low-rank adapters, five backdoor attacks, three backdoor tasks, and \(2{,}495\) test samples per model. We use the original defense parameters for all baselines. Detection time is the main recurring cost after deployment, directly impacting latency. Among defenses that inspect the victim model or adapter to detect abnormal behavior, \method{} has the lowest average detection cost, \(29.69\) ms per sample, compared with \(58.82\) ms for ConfGuard and \(162.83\) ms for BEAT. This is because \method{} requires only a single forward pass, whereas ConfGuard monitors confidence during generation and BEAT performs multiple generations. ONION is faster, averaging \(13.53\) ms per sample, but it directly scans input text, leading to substantially weaker attack rejection as shown in earlier sections.

Setup time is a one-time pre-deployment cost. \method{} takes \(11.98\) seconds on average to compute the rejection threshold and select LoRA insertion sites to monitor. This is comparable to ONION's \(8.72\)-second threshold calibration and substantially lower than BEAT's \(67.46\)-second setup time, which includes threshold calibration and safety probe selection. ConfGuard has no setup cost because it uses a fixed confidence-window rule rather than dynamically selecting parameters.

\section{Limitations and Future Work}
\label{sec:limitations}

First, this work focuses on autoregressive large language models based on Transformer architectures, the predominant design underlying chatbots and code assistants~\cite{qorib2024decoder,grattafiori2024llama}. Future work can extend the proposed backdoor detection framework to emerging non-Transformer architectures, such as state-space models~\cite{gu2023mamba}, and to multimodal models~\cite{liu2023visual}.

Second, \method{} rejects suspicious inputs rather than discarding an entire adapter, thereby reducing the cost of false positives. Another line of research focuses on sanitizing inputs by identifying and removing backdoor triggers before processing them~\cite{zhou2025survey}. Future work can extend \method{} to support input sanitization. One possible approach is to iteratively mask candidate tokens while monitoring the resulting decrease in the activation-spike score. Nevertheless, input sanitization may not always preserve the meaning of the original request. For example, if ``Thanksgiving'' is the trigger in the user query ``What day is Thanksgiving?'', removing it produces an incomplete request.

Third, \method{} addresses a scenario where the defender downloads third-party adapters and maintains white-box access to intermediate activations. In contrast, an end-user querying a black-box API can only send prompts and receive generated text. From such a black-box perspective, the base model and adapter function as a single merged system, causing adapter-specific backdoor signals to be obscured. As empirically demonstrated in earlier sections, adapter-agnostic defenses underperform significantly. Importantly, while black-box API clients lack visibility into internal adapter channels, the platform operator hosting the API possesses full access and can deploy \method{} directly at inference time to safeguard the pipeline.

\section{Ethical Considerations}
\label{sec:ethics}

This work addresses security vulnerabilities in third-party low-rank adapter (LoRA) ecosystems with a focus on defense. Adversaries might attempt to design sophisticated attacks based on our findings, such as adding activation-smoothing penalties during backdoor training. Nevertheless, we believe that the benefits of open research outweigh these risks. Publicly identifying vulnerabilities helps secure production systems before bad actors exploit them \cite{liu2025vulnerability}. Furthermore, all evaluations in this paper were conducted using publicly available, peer-reviewed benchmarks, avoiding the creation of new malicious models, adapters, or datasets.

\section{Conclusion}
\label{sec:Conclusion}

The integrity of third-party LoRA adapters is a growing concern for LLM deployment \cite{liu2026comprehensive}. Existing backdoor defenses either ignore adapter-specific structure by merging adapters with base models, which can dilute adapter-specific backdoor signals, or operate at a higher level by relying on classifier-based rejection of entire adapters. This paper addresses this gap by introducing \method{}, an inference-time backdoor detector designed specifically for low-rank adapters. Before deployment, \method{} selects a small subset of stable LoRA insertion sites, using \(5\%\) by default. During inference, it monitors LoRA down-projection bottleneck activations at these sites and rejects anomalous inputs that exhibit highly concentrated activation spikes. We empirically demonstrate that monitoring this subset substantially outperforms monitoring all LoRA insertion sites. Across a standard LLM backdoor benchmark with diverse models, attacks, and tasks, \method{} achieves \(98.49\%\) attack rejection, outperforming existing baselines without a significant trade-off on clean inputs.

\section*{GenAI Usage Disclosure}

Generative AI was used to improve grammar, assist with LaTeX formatting, and debug code. The authors reviewed all resulting changes for correctness.

\bibliographystyle{ACM-Reference-Format}
\bibliography{references}

\end{document}